\documentclass[prd,twocolumn,showpacs,superscriptaddress,preprintnumbers,
amsmath,amssymb]{revtex4-1}

\usepackage[colorlinks=true,urlcolor=blue,linkcolor=blue,citecolor=blue]{hyperref}
\usepackage[normalem]{ulem}
\usepackage[capitalize]{cleveref}
 
\usepackage{comment}
\usepackage{flushend}
\usepackage[T1]{fontenc} 
\usepackage{mathtools}
\usepackage{xcolor}
\usepackage{graphicx}
\usepackage{dcolumn}
\usepackage{bm}
\usepackage{multirow}
\usepackage{slashed}

\usepackage[export]{adjustbox}

\newcommand{\Beq}{\begin{equation}\begin{aligned}}
\newcommand{\Eeq}{\end{aligned}\end{equation}}

\newcommand{\GW}{\mathrm{GW}}

\begin{document}

\preprint{IPMU23-0036, YITP-23-124, KEK-QUP-2023-0025, KEK-TH-2560, KEK-Cosmo-0328}

\title{Axion-like Universal Gravitational Wave Interpretation of Pulsar Timing Array Data
}

\author{Kaloian D. Lozanov}
\email{kaloian.lozanov@ipmu.jp}
\affiliation{Kavli Institute for the Physics and Mathematics of the Universe (WPI), UTIAS,
The University of Tokyo, Kashiwa, Chiba 277-8583, Japan}

\author{Shi Pi}
\email{shi.pi@itp.ac.cn}
\affiliation{CAS Key Laboratory of Theoretical Physics, Institute of Theoretical Physics, Chinese Academy of Sciences, Beijing 100190, China}
\affiliation{Center for High Energy Physics, Peking University, Beijing 100871, China}
\affiliation{Kavli Institute for the Physics and Mathematics of the Universe (WPI), UTIAS,
The University of Tokyo, Kashiwa, Chiba 277-8583, Japan}

\author{Misao Sasaki}
\email{misao.sasaki@ipmu.jp}
\affiliation{Kavli Institute for the Physics and Mathematics of the Universe (WPI), UTIAS,
The University of Tokyo, Kashiwa, Chiba 277-8583, Japan}
\affiliation{Center for Gravitational Physics and Quantum Information, Yukawa Institute for Theoretical Physics,
Kyoto University, Kyoto 606-8502, Japan}
\affiliation{Leung Center for Cosmology and Particle Astrophysics, National Taiwan
University, Taipei 10617}

\author{Volodymyr Takhistov}
\email{vtakhist@post.kek.jp, *corresponding}
\affiliation{International Center for Quantum-field Measurement Systems for Studies of the Universe and Particles (QUP, WPI),
High Energy Accelerator Research Organization (KEK), Oho 1-1, Tsukuba, Ibaraki 305-0801, Japan}
\affiliation{Theory Center, Institute of Particle and Nuclear Studies (IPNS), High Energy Accelerator Research Organization (KEK), Tsukuba 305-0801, Japan
}
\affiliation{Graduate University for Advanced Studies (SOKENDAI), \\
1-1 Oho, Tsukuba, Ibaraki 305-0801, Japan}
\affiliation{Kavli Institute for the Physics and Mathematics of the Universe (WPI), UTIAS,
The University of Tokyo, Kashiwa, Chiba 277-8583, Japan}

\author{Ao Wang}
\email{wangao@itp.ac.cn}
\affiliation{CAS Key Laboratory of Theoretical Physics, Institute of Theoretical Physics, Chinese Academy of Sciences, Beijing 100190, China}
\affiliation{School of Physical Sciences, University of Chinese Academy of Sciences, Beijing 100049, China}
 
\date{\today}

\begin{abstract}
Formation of cosmological solitons is generically accompanied by production of gravitational waves (GWs), with a universal GW background expected at frequency scales below that of non-linear dynamics. Beginning with a general phenomenological description of GWs associated with soliton formation, we demonstrate that universal GW background from axion-like particle (ALP) solitonic oscillons provides a viable interpretation to the recent NANOGrav 15 year pulsar timing array data, which does not suffer from the overproduction of primordial black holes.
We show that pulsar timing array data displays preference for models where formed solitons do not strongly interact or cluster. Coincidence observations with Nancy Roman telescope will allow to discriminate between distinct scenarios of cosmological solitons. 
\end{abstract}

\maketitle  

\section{Introduction}
Cosmological non-perturbative solitonic defects (e.g. oscillons, Q-balls, monopoles, strings) can readily arise in various theories due to unique extreme conditions of the early Universe~\cite{Vilenkin:2000jqa,Shnir:2018yzp}. Among the prominent mechanism leading to production of cosmological solitons are the Kibble-Zurek mechanism associated with symmetry breaking~\cite{Kibble:1976sj,Zurek:1985qw} or instabilities and fragmentation associated with self-interactions of quantum fields (e.g.~\cite{Amin:2011hj,Cotner:2019ykd}).

A plethora of phenomena have been associated with such cosmological solitons. For example, fragmentation of 
real scalar fields into oscillons frequently appears in theories of inflation favored by observational data~\cite{Amin:2011hj,Lozanov:2017hjm}. Fragmentation of a complex scalar fields into Q-balls is a common feature of theories seeking to explain the observed baryon-antibaryon asymmetry of the Universe (see, e.g.,~\cite{Allahverdi:2012ju}). Furthermore, this has been also linked to formation of primordial black holes~\cite{Cotner:2016cvr,Cotner:2018vug,Cotner:2019ykd}.  
Formation of solitons is generically accompanied by production of stochastic gravitational wave (GW) background (e.g.~\cite{Lozanov:2019ylm,Caprini:2018mtu}). 
 Aside prominent GW spectrum peak at frequencies associated with scales of non-linearities when solitons form, recently it was observed that soliton formation is expected to be accompanied by universal GW (UGW) background contributions at lower frequencies~\cite{Lozanov:2023aez,Lozanov:2023knf}.

Recently, Pulsar Timing Array (PTA) collaborations NANOGrav~\cite{NANOGrav:2023gor,NANOGrav:2023hde}, EPTA combined with InPTA~\cite{EPTA:2023fyk,EPTA:2023sfo,EPTA:2023xxk}, PPTA\,\cite{Zic:2023gta,Reardon:2023gzh,Reardon:2023zen} as well as CPTA\,\cite{Xu:2023wog} have reported evidence for the detection of stochastic GW background signal around nHz frequency range. The origin of inferred signal has not been definitively established. While the signal is broadly consistent with expected incoherent superposition of the GWs emitted from a population of merging supermassive black hole (SMBH) binaries distributed throughout the Universe, interpretation based on minimal models of SMBH binary systems may require atypical parameters \cite{NANOGrav:2023hfp,EPTA:2023xxk} leading to alternative scenarios~\cite{NANOGrav:2023hfp,Ellis:2023dgf,Ghoshal:2023fhh,Shen:2023pan,Broadhurst:2023tus,Bi:2023tib,Zhang:2023lzt} such as including a stronger role of environmental effects.
The signal is also consistent with a cosmological origin and contribution of new physics (see e.g.~\cite{NANOGrav:2023hvm, EPTA:2023xxk,Ellis:2023oxs}), including 
cosmic string networks \,\cite{Ellis:2023tsl,Kitajima:2023vre,Wang:2023len,Lazarides:2023ksx,Eichhorn:2023gat,Chowdhury:2023opo,Servant:2023mwt,Antusch:2023zjk,Yamada:2023thl,Ge:2023rce,Basilakos:2023xof}, phase transitions~\,\cite{Addazi:2023jvg,Bai:2023cqj,Megias:2023kiy,Han:2023olf,Zu:2023olm,Megias:2023kiy,Ghosh:2023aum,Xiao:2023dbb,Li:2023bxy,DiBari:2023upq,Cruz:2023lnq,Gouttenoire:2023bqy,Ahmadvand:2023lpp,An:2023jxf,Wang:2023bbc}, domain walls~\,\cite{Kitajima:2023cek,Guo:2023hyp,Blasi:2023sej,Gouttenoire:2023ftk,Barman:2023fad,Lu:2023mcz,Li:2023tdx,Du:2023qvj,Babichev:2023pbf,Gelmini:2023kvo,Zhang:2023nrs}, primordial fluctuations~\,\cite{Franciolini:2023pbf,Choudhury:2023hfm,Vagnozzi:2023lwo,Franciolini:2023wjm,Inomata:2023zup,Cai:2023dls,Wang:2023ost,Ebadi:2023xhq,Gouttenoire:2023nzr,Liu:2023ymk,Abe:2023yrw,Unal:2023srk,Yi:2023mbm,Firouzjahi:2023lzg,Salvio:2023ynn,You:2023rmn,Bari:2023rcw,Ye:2023xyr,HosseiniMansoori:2023mqh,Cheung:2023ihl,Balaji:2023ehk,Jin:2023wri,Bousder:2023ida,Das:2023nmm,Zhao:2023joc,Ben-Dayan:2023lwd,Jiang:2023gfe,Liu:2023pau,Yi:2023tdk,Frosina:2023nxu,Bhaumik:2023wmw,Domenech:2021and}, audible axions\,\cite{Figueroa:2023zhu,Geller:2023shn} as well as other scenarios\,\cite{Depta:2023qst,Yang:2023aak,Li:2023yaj,Lambiase:2023pxd,Borah:2023sbc,Datta:2023vbs,Murai:2023gkv,Niu:2023bsr,Choudhury:2023kam,Cannizzaro:2023mgc,Zhu:2023lbf,Aghaie:2023lan,He:2023ado}. Some related interpretations have also been suggested for previous data sets of NANOGrav PTA~(e.g.~\cite{Domenech:2020ers,Lu:2022yuc,Sugiyama:2020roc,Sakharov:2021dim}), among others. Further explorations are necessary to establish the origin of signal and its implications for astrophysics and new physics.

In this work, we comprehensively analyze GWs (especially UGWs) associated with soliton formation and explore their compatibility with PTA signals, focusing on NANOGrav 15 year data set. We also discuss the ability of PTA observations to discriminate between models of soliton formation.
\vspace{-1em}

\section{Gravitational Waves from cosmological solitons}

\subsection{Soliton formation}

In order to discuss GWs generated by solitons in the early Universe, we consider the following general cosmological scenario. We assume the solitons form in the epoch of radiation domination, before the onset of Big Bang nucleosynthesis (BBN). The solitons are taken to comprise a subdominant spectator field, whose contribution to the Universe's energy budget is small during the epochs of inflation and radiation domination after it. The spectator field interacts negligibly with the radiation. 

We assume that due to significant field self-interactions, the spectator field forms solitonic objects at time $\tau_i$, such as through field fragmentation mechanism, at a characteristic object wavenumber $k_{\rm sol}$. The typical soliton separation is then given by the nonlinear scale $k_\mathrm{nl}=0.1k_{\rm sol}$ ~\cite{Amin:2011hj}. Since on scales $k<k_\mathrm{nl}$ the formation events are expected to be independent, the soliton density field obeys Poisson statistics. On scales $k>k_\mathrm{nl}$, the soliton density field peaks at $k_{\rm sol}$.

Once formed, the solitons are taken to behave as pressureless dust. Hence, their energy density decays with scale factor $\propto a^{-3}$, more slowly than radiation ($\propto a^{-4}$). Eventually, at time $\tau_{\rm eq}\equiv1/k_{\rm eq}$, solitons start to dominate. Note that this is distinct time from standard cosmological matter-radiation equality. Here, for simplified analysis, we consider that at the moment $\tau_{\rm eq}$ solitons decay into radiation, however in general these timescales can be distinct.

\subsection{Gravitational wave spectrum}
\label{ssec:gravspec}

The scenario outlined above could give rise to distinct classes of GWs, which allow us to formulate general phenomenological description that we will employ to fit the GW data.
First, the process of soliton formation generates high-frequency GWs on scales $k>k_\mathrm{nl}$, with a characteristic peak at $k_\mathrm{sol}$~\cite{Amin:2014eta}. For simplicity, we assume that the solitonic objects are spherically symmetric (e.g. Q-balls, oscillons, etc.), so that once formed they do not emit significant amount of GWs anymore due to the absence of additional structure and quadrupole moment. We note that our analysis can be generalized to include such additional GW effects.

The nonlinear dynamics during soliton formation could yield a sizeable transverse traceless component of the energy momentum tensor of the spectator field. This sources GWs with a peak frequency determined by $k_\mathrm{sol}$, and an amplitude of \cite{Lozanov:2019ylm}
\Beq
\label{eq:B}
B\sim \Omega_{\rm r}(\delta^{TT})^2\left(\frac{\mathcal{H}_i}{k_{\rm sol}}\right)^{2}\,,
\Eeq
where $\mathcal{H} = a H$ with $H_i$ being the Hubble parameter at soliton formation time $\tau_i$, and $\delta^{TT}$ is the fraction of the critical energy of the Universe stored in the transverse-traceless part of the energy momentum tensor of the soliton fluid and the parameter is calculated at the time of soliton formation. Here, $\Omega_{\rm r}\sim10^{-5}$ is the fraction of the critical energy of the Universe today stored in relativistic degrees of freedom.  

Hence, taking the shape of the peaked GW spectrum to follow approximately log-normal distribution, 
motivated by lattice simulations~(e.g.~\cite{Lozanov:2019ylm}),
we can parameterize these GW contributions as
\Beq
\label{eq:GWsSol}
\Omega_{\rm 
 GW,0}^{\rm sol}\Big(\dfrac{f_0}{{\rm Hz}}\Big)=B\exp\Big[- \ln^2\Big(\dfrac{f_0}{10~ C}\Big) \Big]~,
\Eeq
where the scale of the central frequency can be obtained from cosmological conditions as~\cite{Lozanov:2019ylm}
\Beq
\label{eq:C}
C\simeq\frac{k_{\rm sol}}{\mathcal{H}_i}\frac{\rho_{i}^{1/4}}{m_{\rm pl}}\times10^9~{\rm Hz}~,
\Eeq
where $m_{\rm pl}$ is the Planck mass, and $\rho_{i}$ is the energy density of the Universe at the time of soliton formation, which is approximately energy density of thermal bath.

In addition to the processes outlined above, for causal soliton formation one can generically expect generation of lower-frequency UGWs~\cite{Lozanov:2023aez,Lozanov:2023knf}. This can be understood by noting that since the soliton formation events are uncorrelated on large scales, the soliton density field obeys Poissonian statistics on $k<k_\mathrm{nl}$. These variations in the density of the effective soliton fluid on large scales appear as isocurvature perturbations while the solitons are still energetically subdominant with respect to the radiation. The isocurvature perturbations can in turn source curvature perturbations, which generate UGWs at second order in scalar perturbations on $k<k_\mathrm{nl}$. 

We parameterize contributions from UGWs~\cite{Lozanov:2023aez} sourced by the large-scale perturbations in the density field of the solitons as
\Beq
\label{eq:UGWs}
\Omega_{\rm GW,0}^{\rm UGW}\Big(\dfrac{f_0}{\rm Hz}\Big)=
\frac{A}{\Big(\dfrac{10^2 f_0}{C}\Big)^{-3}+\Big(\dfrac{5 f_0}{C}\Big)^{-1.8}+F\Big(\dfrac{ f_0}{C}\Big)+1}\,,
\Eeq
where
\Beq
\label{eq:A}
A\sim 10^{-1}\Omega_{\rm r}\left(\frac{k_{\rm eq}}{k_\mathrm{nl}}\right)^4\,.
\Eeq
The parametrization captures general features of expected of oscillon UGWs~\cite{Lozanov:2023aez}, in particular related to peaked behavior with extended tail associated with induced GWs from isocurvature power spectrum contributions on quasi-linear scales.
The function $F(f_0/C)$ is chosen such that it vanishes for frequencies sourced by perturbations on scales $k<k_\mathrm{nl}$, but it suppresses the $\Omega_{\rm GW,0}^{\rm UGW}$ on the scales $k>k_\mathrm{nl}$ relevant to the GWs at higher frequencies that are associated with soliton formation. For concreteness, we choose $F(f_0/C)=(f_0/10~C)^{10}$. Our analysis can be easily extended to other choices of phenomenological formulae.

In discussion of UGWs above, we have assumed that the objects are uncorrelated and follow Poissonian statistics. However, non-Poissonian distribution can naturally arise due to soliton interactions~\cite{Lozanov:2023knf}. In the generic case of gravitational clustering, the isocurvature power spectrum of solitons is expected to follow $\mathcal{P}_S \propto k^2$ instead of $\mathcal{P}_S \propto k^3$. The resulting GW spectrum for lower frequency ranges flattens out~\cite{Lozanov:2023knf}. This can be accounted for by a phenomenological parametrization in Eq.~\eqref{eq:UGWs} by modifying the exponential power of the terms with $f_0$ in the denominator.

\begin{figure}[t] \centering 
\includegraphics[width=0.48\textwidth]
{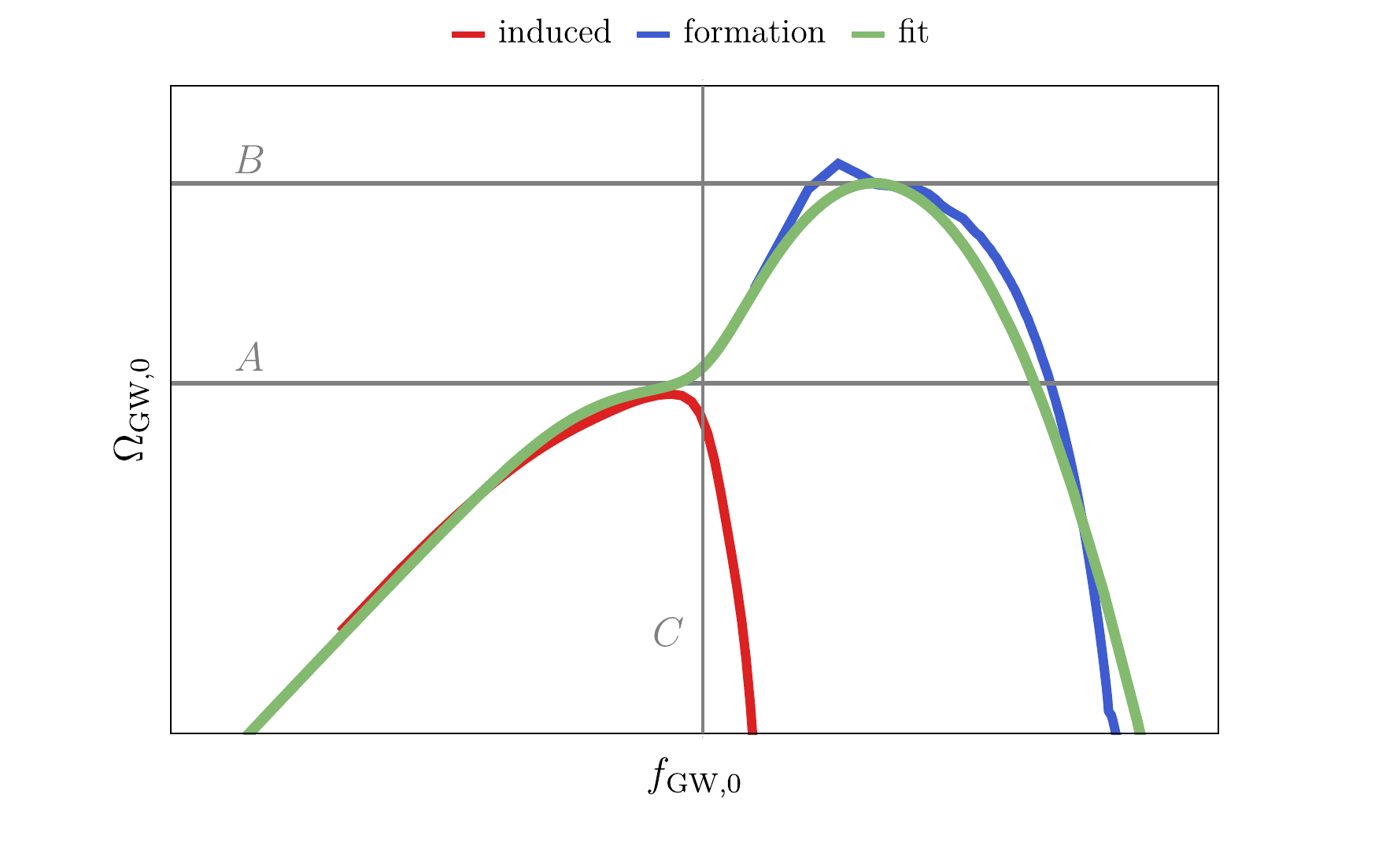}
\caption{Fit of phenomenological soliton GW formula to ALP oscillon GW spectrum. The parameter $A$ defined in Eq. \eqref{eq:A} determines the amplitude of the induced GW signal. $B$ defined in Eq. \eqref{eq:B} sets the amplitude of the GWs from formation. $C$ defined in \eqref{eq:C} fixes the frequency of the overall GW signal. Note the ratio of the peak frequency of the induced GWs signal and the one from formation is fixed. For more details see text.
} 
\label{fig:schematic}
\end{figure}

Combining Eq.~\eqref{eq:GWsSol} and Eq.~\eqref{eq:UGWs}, we put forth the following phenomenological formula for GWs associated with soliton production
\Beq
\Omega_{\rm GW,0}(A,B,C,f_0) =~\Omega_{\rm GW,0}^{\rm sol} (B,C,f_0) +\Omega_{\rm GW,0}^{\rm UGW}(A,C,f_0)~.
\label{eq:Gwstot}
\Eeq
This description relies on three phenomenological parameters: $A$, $B$ and $C$. The meanings of these parameters are illustrated in Fig.~\ref{fig:schematic}.

We note that additional contributions to soliton GWs can arise from their decay, as demonstrated for oscillons in Ref.~\cite{Lozanov:2022yoy}. However, these contributions depend on additional model considerations and here we are primarily interested in analyzing the effects of UGWs.

\subsection{Axion-like particle (ALP) oscillons}
\label{sec:ALPoscillons}

As a concrete realisation of our scenario, we consider the formation of oscillons by an axion-like particle (ALP) field, following analysis of Ref.~\cite{Lozanov:2023aez}. The ALP field constitutes a light spectator during inflation in early Universe, with a sub-Hubble mass and sub-dominant contribution to energy density. Some time after the end of inflation, during radiation domination era, the mass of the ALP field becomes comparable to the Hubble scale with $m\sim H_i$. The ALP background begins to oscillate around the minimum of its potential 
\begin{equation}
    V=m^2F^2\Big[1-\cos\Big(\dfrac{\phi}{F}\Big)\Big]~,
\end{equation}
with axion decay constant $F\sim m_\mathrm{pl}\sqrt{k_\mathrm{eq}/\mathcal{H}_i}$~\cite{Lozanov:2023aez}. Its self-interactions lead to resonant instabilities in the small-scale perturbations of the ALP. The perturbations rapidly back-react and the ALP field becomes non-linear. The non-linear dynamics lead to field fragmentation into lumps and formation of solitonic oscillons. We also assume that before the onset of BBN the oscillons decay into Standard Model (SM) radiation, due to a coupling between the ALP field and, e.g., SM photons \cite{Hertzberg:2018zte}.

\subsection{Restrictions from CMB}
\label{ssec:cmb}

The energy density contributed by GWs produced in the early Universe also drives the expansion of Universe. The maximum possible contribution from GWs is restricted by the difference between the observed Hubble expansion rate and its Standard Model prediction. These effects can be parameterized by the deviation between the number of effective relativistic neutrino species and that of Standard Model, $\Delta N_{\rm eff}$. Hence, GWs produced during the soliton formation are bounded by~\cite{Caprini:2018mtu}
\begin{align}
    \Omega_{\GW,0}h^2 &=\int_{f_\mathrm{min}}^{\infty} \Omega_{\GW,0}(f) h^2\frac{d f}{f}\notag \\ 
    &< \Omega_{\gamma,0} h^2\left(\frac{g_S(T_0)}{g_S(T)}\right)^{4/3}\frac{7}{8}\Delta N_{\rm eff},
\end{align}
where $T$ represents the background temperature of the relevant event, $f_\mathrm{min}$ is approximately equal to the inverse of the age of the Universe at that moment, and $\Omega_\gamma$ is the cosmological parameter of photons. The subscript ``${}_0$'' denotes quantities at present time.

Since the temperatures of BBN and cosmic microwave background (CMB) are known, the observable quantities, i.e. the abundance of light elements from BBN as well as the location of acoustic peaks on CMB, will be modified if the energy densities during those events are altered. In order not to be distracted by details, we consider a robust upper bound of $N_{\rm eff} <3.28$ as obtained at the 95\% confidence interval by the Planck 2018 data analysis~\cite{Planck:2018vyg}. With $g_S(T_{\mathrm{CMB}})=g_S(T_{0})=3.91$, $\Omega_{\gamma,0} h^2=2.47\times10^{-5}$ and $h=0.683$, we obtain $\Omega_{\GW,0}<1.08\times10^{-5}$, although a slightly stringent bound can be obtained at $T_\mathrm{BBN}$ as $g_S(T_{\mathrm{CMB}})<g_S(T_{\mathrm{BBN}})$. Our discussion  can be readily extended to account for additional experiments.

\subsection{Possible black hole production}

We note that additional constraints can arise from overproduction of primordial black holes~(PBHs)~\cite{Cotner:2018vug,Cotner:2019ykd}. While we leave detailed analysis of these aspects for future work, as we argue below, PBH formation will not jeopardize our scenarios.

Once solitons form, they can cluster due to their interactions (e.g., gravitational attraction). A fraction of the soliton clusters could be compact enough to form (collapse into) PBHs. PBHs with mass at formation $\gtrsim 10^{15}g$ can survive Hawking evaporation until present day, of which the overproduction is dangerous as it contradicts observational constraints. Such PBHs cannot evaporate through soliton decay channels. However, in our soliton formation scenarios, such PBH formation and associated early matter domination epoch are improbable. The lifetime of our solitons is generically too short for significant clustering to take place, as there are only a few $e$-folds of expansion between formation and decay. During this period only a negligible fraction of the solitons will form compact object clusters. Hence, we do not expect that exponentially small fraction of solitons forming PBHs can lead to dangerous matter domination around e.g. era of BBN. As our solitons are produced in time near the BBN epoch, corresponding possible PBHs arising from soliton collapse are not expected to come to dominate the radiation in the Universe in the intervening period.

\begin{figure*}[t] \centering 
\includegraphics[width=0.48\textwidth,valign=c]
{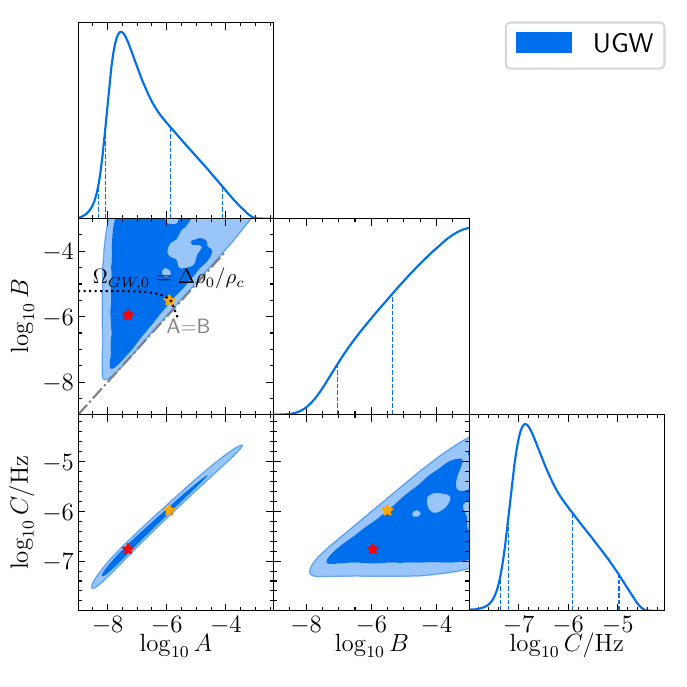}
\begin{minipage}{0.5\linewidth}
\includegraphics[trim={0 0 0cm 0cm},clip, 
width=1\textwidth,valign=c]
{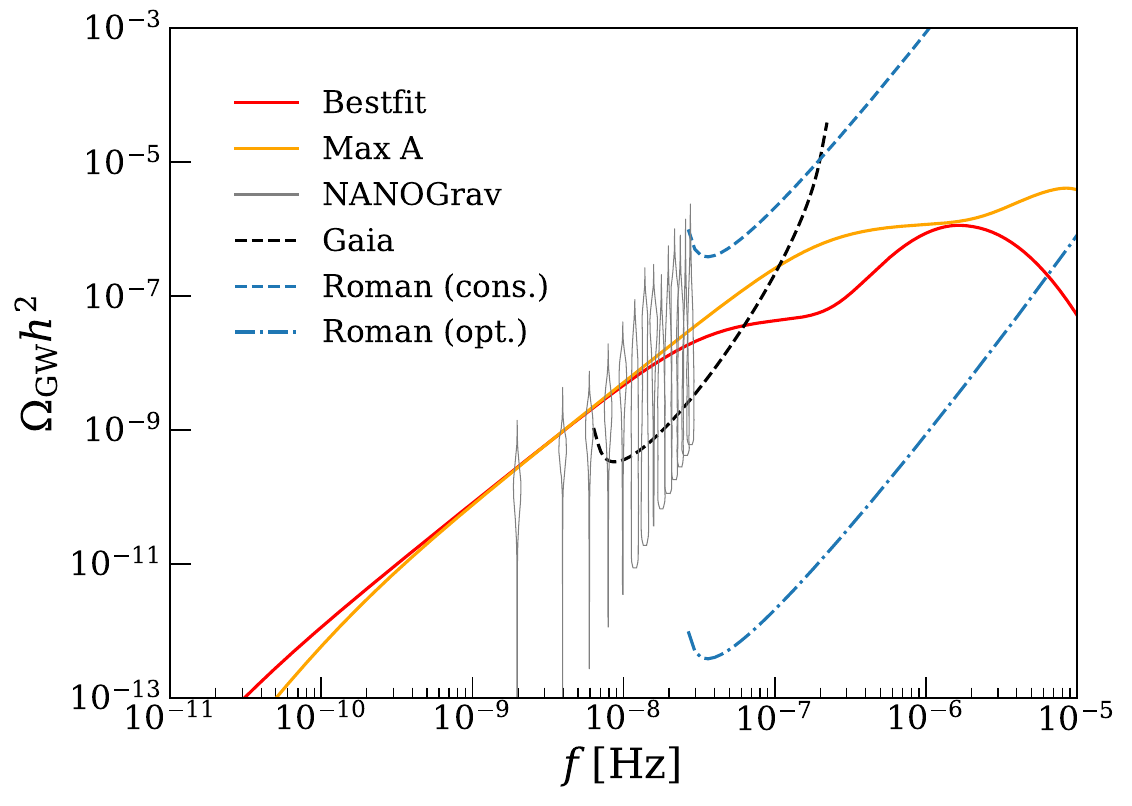}
\end{minipage}
\caption{\textbf{[Left]} Reconstructed posterior distribution for the parameters of soliton formation GW scenario. After removing the samplers violating the CMB bound on $N_{\rm eff}$, we are able to find the best-fit point: $[\log_{10}A,\log_{10}B,\log_{10}C]=$[-7.3, -6.0, -6.8] in remained allowed parameter space, which is marked as a red star. And the yellow star marks the point with the maximal amplitude of $A$ in the 1-$\sigma$ which doesn't violate the CMB bound, $[\log_{10}A,\log_{10}B,\log_{10}C]$=[-5.9, -5.5, -6.0].
\textbf{[Right]} The red line shows the energy spectrum of the best-fit parameters. The yellow line is the spectra of the Poissonian tail with the maximal amplitude of $A$ and doesn't violate the CMB bound. The grey lines are the violin plots for the first 14 frequency bins of NANOGrav 15-year data.
} 
\label{fig:GWPsFIT}
\end{figure*}

\section{Pulsar Timing Array Data Fit}

\subsection{Fit of gravitational waves from solitons}

For the fit of soliton GWs characterized by Eq.~\eqref{eq:Gwstot} to PTA data, specifically focusing on NANOGrav, we employ Bayesian fit with Bayes estimator for our parameter space computed with \href{https://andrea-mitridate.github.io/PTArcade/}{\textsc{PTArcade}} package in Ceffyl mode~\cite{lamb2023need}. The parameters and their priors for the Monte Carlo Markov Chain (MCMC) sampling and the 95\% highest probability density interval (HPDI) are displayed in Tab.~\ref{tab:bayesfitUGWs}. Since the amplitude of the UGWs arising from Poisson distribution tail is physically limited to be smaller than the main peak of GWs from soliton formation, we considered $\log (A/B)$ as one of our MCMC parameters with a prior distribution restricted to be less than 1. However, the results of the fitting procedure are still expressed by the parameters used in Eq.~\eqref{eq:Gwstot}. 

Considering UGWs to be the main contribution to the PTA time residual correlation signatures as we discuss above, the frequency $f_\mathrm{nl}$ is expected to lie approximately in the region of the PTA working frequency $f_{\mathrm{PTA}}$. In order to avoid restrictions imposed on GW energy density by CMB $N_{\rm eff}$ (see Sec.~\ref{ssec:cmb}),
we consider a conservative cut-off for $\log_{10}B$. After integrating over Eq.~\eqref{eq:Gwstot}, the total GW contribution of our scenario is given by
\begin{equation}
    \Omega_{\GW,0} \simeq 3.9 A + \sqrt{\pi} B~,
\end{equation}
where the coefficient of A is numerically given and the $f_\mathrm{min}$ is approximated to 0 since the PTA working frequencies $f_{\mathrm{PTA}}$ is much larger than $f_{\mathrm{CMB}}$. 
Hence, to discuss model parameters most preferred by NANOGrav data we remove the points violating the GW energy density bound due to CMB.

\begin{table}
\begin{tabular}{l|c|c|c}
\hline
\hline
         & $\log_{10}(A/B)$ & $\log_{10}B$       & $\log_{10}C$       \\ \hline 
Prior    & Uniform{[}-5,0{]}      & Uniform{[}-9,-4{]} & Uniform{[}-8,-4{]} \\ \hline
         & $\log_{10}A$ & $\log_{10}B$       & $\log_{10}C$       \\ \hline
95\%HPDI & {[}-8.33,-4.12{]}      & {[}-7.36,NaN{]}    & {[}-7.36,4.97{[}    \\ \hline \hline
\end{tabular}
\caption{The fit priors follow uniform distribution and the corresponding values are written as NaN if the HPDI bound is equal to the prior.}
\label{tab:bayesfitUGWs}
\end{table}

Performing the fit, the posterior distributions are shown in the left panel of ~Fig.~\ref{fig:GWPsFIT}, with the best-fit point (yellow dot). The right panel of Fig.~\ref{fig:GWPsFIT} displays the GW energy spectra assuming best fit parameters for Eq.~\eqref{eq:Gwstot}. We compare them with the first 14 frequency ``violin plot'' bins of the NANOGrav 15-year data set as in Ref.~\cite{NANOGrav:2023gor}, to reduce correlations with excess noise at higher frequencies. Further, we display (``Max A'') the GW spectrum fit
with the maximum possible value of A within 1-$\sigma$ region which does not violate the CMB $N_{\rm eff}$ bound. Since our UGW fit mainly relies on the term in denominator of Eq.~\eqref{eq:UGWs} that is $(5 f_0/C)^{-1.8}$ for interpretation of the lower frequency GW signal, we favor values of parameter $C$ that will enable $f_\mathrm{PTA}$ to be located within the interval that $(5 f_0/C)$-term dominates.

Astrometry offers a complementary approach to PTAs for observing the SGWB. With astrometric techniques, apparent changes in stellar positions on the sky due to passing GWs can be tracked. The distinctive correlation patterns in these positional shifts allow for the inference of GW background properties. Using relative astrometry, NASA's upcoming Nancy Grace Roman Space Telescope’s Galactic Bulge Time Domain Survey (GBTD)~\cite{WFIRST:2019}, which is expected to observe approximately $10^8$ targets imaged around $\sim 4 \times 10^5$ times toward the Galactic Center, can provide valuable data for high sensitivity GWs studies~\cite{Wang:2022sxn,Pardo:2023cag}. Roman is well positioned for frequent and numerous exposures of the same target region. It is expected to have 15-minute observing cadence
with six 72-day observational seasons spread out over the nominal 5-year mission time. One can infer~\cite{Wang:2022sxn,Pardo:2023cag} that Roman will be sensitive to GWs in the $7.7 \times 10^{-8}$~Hz~$ < f < $~$5.6 \times 10^{-4}$~Hz range. Roman is poised to probe stochastic GW background with an estimated sensitivity of SNR $\sim 70$, considering a signal of around the range $h_c(f) = 2.8 \times 10^{-15}({\rm f/1 yr})^{-2/3}$.
Hence, Roman will be able to not only cross-check PTA signals over certain frequency ranges but 
importantly also probe microHz frequencies, allowing to further test distinct GW background sources as well as bridge the frequency-band gap between PTAs and upcoming space-based interferometer experiment LISA~\cite{LISA:2017pwj}.

On the other hand, Gaia astrometry survey is already underway. While full survey datasets have yet to be released, considering cleanest datasets of Gaia DR3 data release with $\mathcal{O}(10^6)$ observed sources allows to restrict GW background signal to be~\cite{Jaraba:2023djs} $\Omega_{\rm GW}h^2 \lesssim 8.7 \times 10^{-2}$ over the range $4.2 \times 10^{-18}$~Hz~$ < f < $~$1.1 \times 10^{-8}$~Hz. On the right panel of Fig.~\ref{fig:GWPsFIT} we display the potential ultimate sensitivity reach estimates of Roman and Gaia~\cite{Wang:2022sxn}, highlighting that such observations will be able to further restrict the parameter space of our scenario and discriminate between GWs originating from soliton models by considering coincidence with PTA signatures.

From results of phenomenological fit to NANOGrav data, we can now obtain corresponding underlying corresponding ALP oscillon model parameters.
From the Bayesian analysis best fit parameters of $A=10^{-7.3}$, $B=10^{-6}$, $C=10^{-6.8}$, using Eq.~\eqref{eq:B}, Eq.~\eqref{eq:C}, Eq.~\eqref{eq:A} and taking dimensionless energy-momentum tensor $\delta^{TT}\sim1$, we obtain $k_{\rm eq}/k_\mathrm{nl}\sim0.5$, $k_{\rm sol}/\mathcal{H}_i\sim 3$ and $\rho_i^{1/4}\sim 120$~GeV. From this, it follows that $k_{\rm eq}/\mathcal{H}_i\sim 0.5$. Hence, for the ALP potential parameters we have $m\sim H_i\sim 6\times10^{-6}$~eV and $F\sim m_{\rm pl}\sqrt{k_{\rm eq}/\mathcal{H}_i}\sim 0.7m_{\rm pl}$. This implies that ALPs with a sub-Planck scale decay constant, fragmenting into oscillons shortly before the QCD phase transition, can provide a possible explanation to the observed PTA signal.
Following \cite{Arvanitaki:2019rax} we estimate that in our scenario oscillons can be produced for typical initial misalignments (e.g., $\theta_0=2.5$) within several oscillations of the axion background, if the initial amplitude of adiabatic fluctuations is large enough on the relevant $k$ scales for parameteric resonance , i.e., power spectrum amplitude of order $10^{-3}$ on $k\sim10^{22} {\rm Mpc}^{-1}$. Further, we expect that in our setup oscillon formation can be efficient, with majority of axion particles contributing to oscillons, once gravitational attraction is accounted for. Here, the consideration of efficient coalescence of axions into oscillons is important, as it allows to avoid early matter domination caused by free, stable massive axions. We leave detailed numerical simulation analysis of these effects for future work.

For completeness, we outline a scenario for the decay of the best-fit ALP oscillons to SM degrees of freedom. In order to recover the successful BBN scenario, all of the oscillon energy must be converted into SM particles before the onset of BBN. We can consider the case when oscillons decay to SM photons at the time of radiation-oscillon equality. This can be realised by assuming an ALP-photon interaction $\mathcal{L}_{int}=(1/4)g_{\phi\gamma\gamma}\phi F_{\mu\nu}\tilde{F}_{\mu\nu}$. Since the ALP oscillons can transfer energy into photons exponentially fast, $\propto e^{\mu_{\phi\gamma\gamma}t}$, $\mu_{\phi\gamma\gamma}\sim g_{\phi\gamma\gamma} F m$, as a consequence of parametric resonance due to the oscillating $\phi$ background within the oscillon interior \cite{Hertzberg:2018zte,Amin:2021tnq}, the decay into photons is accomplished if $\mu_{\phi\gamma\gamma}> H$. For the best-fit ALP oscillons, this implies $g_{\phi\gamma\gamma}> 10^{-20}~{\rm GeV}^{-1}$, consistent with experimental bounds on the axion-photon coupling \cite{Batkovic:2021fzr}. We note that even if the values of interest for $g_{\phi\gamma\gamma}$ are unnaturally large, $g_{\phi\gamma\gamma} \gg F^{-1}$, from an effective theory point of view, there are high-energy physics models that can allow for this \cite{Agrawal:2018mkd}. Additional constraints on $g_{\phi\gamma\gamma}$ can become relevant if ALP also plays the role of the inflaton instead of spectator field~(e.g.~\cite{Barnaby:2011vw}).

\subsection{Model comparison}

In order to statistically compare goodness of fit among different physics models, one should calculate the relative Bayes factor among them. However, a simplified approximation, the Bayesian information criterion (BIC), is also widely used, which provides
an alternative criterion for model selection among set of models, 
\begin{equation}
    {\rm BIC}=k\ln(n)-2\ln \mathcal{L}~.
\end{equation}
Here,  $k$ is the model parameter number, $n$ is the sampler number, which is 14 for NANOGrav 15 year date set, and $\ln \mathcal{L}$ is the likelihood for the best-fit point. A lower BIC value generally signifies a better fit.

We first compare our UGW fit of NANOGrav data to the expected GW background from population of merging supermassive black hole (SMBH) binary systems. Here, we assume as benchmark the simplest SMBH binary model, with binary evolution primarily described by GW emission~(e.g.~\cite{NANOGrav:2023hfp}). 
Hence, the GW spectrum contributed by GW emission from the SMBH binaries is~\cite{NANOGrav:2023hvm}
\begin{equation}
    \Omega_{\mathrm{GW}}^{\rm SMBH}(f) = A_{\mathrm{BHB}}^2 \frac{2 \pi^2 f^{(5-\gamma_{\rm BHB})}}{3H_0^2}~, 
\end{equation}
where $A_{\rm BHB}$ is the amplitude and $H_0 = 68.3$ km s$^{-1}$ Mpc$^{-1}$. We fit $A_{\rm BHB}$ but fix the spectral index to be $\gamma_{\rm BHB} = 13/3$ expected from analytic calculations~\cite{Phinney:2001di}.

Further, we make comparison with induced GWs from scalar perturbations at second order (see e.g.~\cite{Baumann:2007zm,Mollerach:2003nq,Ananda:2006af,Kohri:2018awv}) that can arise in variety of theories. The induced GW contribution is given by
\begin{align}
\label{eq:Osigw}
\Omega_{\scriptscriptstyle\rm GW}^{\rm ind}\left(f\right) =&~ \Omega_{\rm r} \left(\frac{g_*\left(f\right)}{g_*^0}\right)\bigg(\frac{g_{*,s}^0}{g_{*,s}\left(f\right)}\bigg)^{4/3} \\
&~\times \int_0^\infty \textrm{d}v \int_{\left|1-v\right|}^{1+v} \textrm{d}u\: \mathcal{K}\left(u,v\right) \mathcal{P}_\mathcal{\scriptscriptstyle R}\left(uk\right) \mathcal{P}_\mathcal{\scriptscriptstyle R}\left(vk\right) \,.\notag
\end{align}
where $\mathcal{K}$ is the integration kernel~(e.g.~\cite{Kohri:2018awv,Pi:2020otn,Domenech:2021ztg}), $\Omega_r/g_{*}^0 \simeq 2.72 \times 10^{-5}$ is the current radiation energy
density per relativistic degree of freedom in units of the
critical density~\cite{Planck:2018vyg},
$g_{*,s}^0 \simeq 3.93$ is the number of effective relativistic degrees of freedom contributing
to the radiation entropy at present,  and $g_{*}(f)$, $g_{*,s}(f)$
are the effective numbers of relativistic degrees of
freedom in the early Universe when GWs with comoving
wavenumber $k$ reenter the Hubble horizon. $k$ is connected to the frequency $f$ we observe today by
$k = 2\pi f$, with the current scale factor $a_0$ to be 1.

For model-independent analysis we do not discuss a particular inflation model setup, but consider two generic cases of power spectra, modeled as Dirac delta function or a Gaussian (lognormal) peak in $k$-space 
\begin{equation}\label{PR(k)}
  \mathcal{P}_{\mathcal{\scriptscriptstyle R}}\left(k\right) =
    \begin{cases}
      A\,\delta\left(\ln k-\ln k_*\right), & \text{Delta};\\
      \\
      \dfrac{A}{\sqrt{2\pi}\,\Delta}\,\exp\left(-\dfrac{\ln^2(k/k_*)}{2\Delta^2}\right), & \text{Gauss.}\\
    \end{cases}       
\end{equation}
where $A$ is the amplitude of the total power, $k_*$ is peak scale 
and $\Delta$ is the width. GWs induced by scalar perturbation with a log-normal power spectrum similar to the lower line of \eqref{PR(k)} (SIGW-Gauss) have been found to be the best-fit model, in the sense that it has the largest Bayes factor among all the cosmological interpretations discussed in Ref. \cite{NANOGrav:2023hvm} by NANOGrav collaboration. Here, for our fit we employ model inputs provided by NANOGrav~\footnote{For details, see \href{https://zenodo.org/record/8084351}{https://zenodo.org/record/8084351}.}, but we will use BIC to compare models. 

In addition to the fit of UGWs described above, we also consider comparison with non-Poissonian distribution originating from gravitational interactions as described in Sec.~\ref{ssec:gravspec}. As demonstrated in Ref.~\cite{Lozanov:2023knf}, the resulting lower frequency UGW tail would significantly flatten out compared to Poissonian distribution case that describes solitons without significant interactions and clustering. Assuming the primary contribution to NANOGrav data originates from UGWs associated with solitons, we can compare the results of UGWs fit to a reference fit of a flat (frequency-invariant) GW spectrum that can be taken to represent soliton models with gravitational interactions and soliton clustering, whose UGW spectrum becomes nearly frequency invariant (see Ref.~\cite{Lozanov:2023knf}).

The results of our model comparisons and the corresponding BIC values are summarized in Tab.~\ref{tab:modelcomparison}.
We note that ALP UGW best fit model has difference of $\Delta \mathrm{BIC}= 0.9$ compared to SIGW-Gauss, and $\Delta \mathrm{BIC}= 1.1$ compared to SIGW-Delta, as well as $\Delta \mathrm{BIC}= -4.7$ compared to SMBHs (assuming fixed power index $\gamma_{\rm BHB}$). Hence, UGW constitutes a competitive cosmological interpretation of the NANOGrav signal data, 
but does not suffer from the PBH overproduction which spoils the SIGW interpretation \cite{NANOGrav:2023hvm,Franciolini:2023pbf,Liu:2023ymk,Firouzjahi:2023xke}. Since UGW fit has $\Delta \mathrm{BIC}= -5.0$ compared to a reference flat GW spectrum, we can interpret this as signifying that in the context of interacting solitons the NANOGrav data shows preference for models without significant soliton interactions and clustering.

\begin{table}
\begin{tabular}{l|c}
\hline
\hline
         & BIC      \\ \hline 
         UGWs (Poisson)           & 122.8      \\
         Flat GW spectrum          & 127.8      \\
         SIGW-Delta             & 121.7      \\
         SIGW-Gauss             & 121.9      \\
         SMBH ($\gamma_{\rm BHB} = 13/3$)            & 127.5      \\
         \hline 
         \hline
\end{tabular}
\caption{Comparison of Bayesian information criterion (BIC) of UGWs from ALP soliton formation (best fit point) to different models fitted to NANOGrav 15 year data set.}
\label{tab:modelcomparison}
\end{table}

\section{Conclusions}

Cosmological solitons arise in broad variety of theories and their formation is accompanied by generation of stochastic GW background. Starting with a general phenomenological description, we demonstrated the ability of PTA data to probe GWs arising from soliton formation. We find that UGWs associated with ALP oscillons can address the origin of the recently observed signal in NANOGrav 15 year data set and provide an improved fit compared to simple model of SMBH GWs. 
While the fit of UGW is slightly worse than that of SIGW, the soliton scenario does not have PBH-overproduction problem, which makes it more competitive.
We showed that PTA data displays preference for models with solitons without significant interactions or clustering. Future coincidence observations by Nancy Roman telescope will allow to further discriminate between different cosmological soliton formation models. Our analysis is general and can be applied to other sources of GWs at different frequencies.

~\newline
\textit{Acknowledgements.}---
This work is supported by the National Key Research and Development Program
of China Grant No. 2021YFC2203004, and by World Premier International
Research Center Initiative (WPI), MEXT, Japan. S.P. is partly supported by Project 12047503 of the National Natural Science Foundation of China. This work is also supported in part
by JSPS Grant-in-Aid for Early-Career Scientists No. JP20K14461 (S.P.); by the JSPS KAKENHI grant Nos. 20H05853 (M.S.), 24K00624 (M.S.) and 23K13109 (V.T.).

\bibliography{ref}

\end{document}